\documentclass[epj]{svjour}
\usepackage{amsfonts}
\usepackage{graphics}

\begin{document}

\title{Lattice QCD with a twisted mass term and a strange quark}
\author{Abdou M. Abdel-Rehim\inst{1}\thanks{\emph{Present address:}
        Department of Physics, Baylor University, Waco, TX, USA 76798-7316},
        Randy Lewis\inst{1}\thanks{Speaker.}, R. M. Woloshyn\inst{2}
        \and Jackson M. S. Wu\inst{2}
}

\institute{Department of Physics, University of Regina, Regina, SK, S4S 0A2
      \and TRIUMF, 4004 Wesbrook Mall, Vancouver, BC, V6T 2A3, Canada}

\date{October 2006}

\abstract{
There are three quarks with masses at or below the characteristic scale of QCD
dynamics: up, down and strange.
However, twisted mass lattice QCD relies on quark doublets.
Various options for including three quark flavors within the twisted mass
approach are explored by studying the kaon masses, both analytically
(through chiral Lagrangians) and numerically (through lattice simulations).
Advantages and disadvantages are revealed for each ``strange and twisted''
option.
} 

\authorrunning{Abdel-Rehim, Lewis, Woloshyn and Wu}
\titlerunning{Lattice QCD with a twisted mass term and a strange quark}

\maketitle

\section{Two flavor twisted mass lattice QCD}

Consider a hypercubic lattice in four-dimensional spacetime, and a two flavor
system:
\begin{equation}
\psi(x) = \left(\begin{array}{c}u(x)\\d(x)\end{array}\right) \,.
\end{equation}
In the $m_u=m_d\equiv m_q$ limit, the fermion action for twisted mass lattice
QCD is
\begin{equation}
S_{\rm fermion} = a^4\sum_x\bar\psi(x)\left[D(r,\omega)+m_q\right]\psi(x) \,,
\end{equation}
with Dirac and Wilson terms,
\begin{eqnarray}
D(r,\omega) &=& \gamma\cdot\nabla+\exp(-i\omega\gamma_5\tau_3) W(r) \,, \\
W(r)\psi(x) &\equiv& M_{cr}(r)\psi(x) - \frac{ar}{2}\Box\,\psi(x) \,.
\end{eqnarray}
As usual, the covariant derivative $\nabla_\mu$ and d'Alembertian $\Box$
contain the gauge field $U_\mu(x) \equiv \exp(iagT^bA_\mu^b(x))$.

Two key features of twisted mass lattice QCD are (a) the existence of a firm
lower bound for the eigenvalues of $D(r,\omega)+m_q$, as long as $m_q$ and
$\omega$ are non-zero\cite{FreGSW} and (b) the absence of $O(a)$ errors in
observable quantities at maximal twist,
$\omega(renormalized)=\pm\pi/2$\cite{FreR1}.
Another significant consequence of $\omega\neq0$ is that parity and flavor
symmetries are broken; they only get restored in the continuum limit.
\vspace{-2mm}

\section{Options for including the strange quark}

Given the phenomenological importance of $u,d,s$ physics, how should
the strange quark be added to the two flavor twisted mass theory of the
previous section?

One option is to leave the strange quark untwisted.
It will not be protected from near-zero modes, but this is not a practical
problem for simulations at the physical strange quark mass.  
There will be $O(a)$ errors in the theory, but one could remove their effects
on observables through addition of a Sheikholeslami-Wohlert term\cite{SheW}.
Kaon operators, containing one twisted and one untwisted fermion, would
appear as parity mixtures, requiring extra effort for the extraction of
continuum physics.

Another option is to twist the strange quark by using the charm quark as its
twisting partner.  To accommodate $m_c\neq m_s$, the action for this ``heavy''
doublet will need one additional term relative to the ``light'' doublet action
used in the previous section for up and down.
The quark mass terms in the twisted basis are
\begin{eqnarray}
{\cal L}&=&\bar\psi_l\left(m_l+i\gamma_5\mu_l\tau_3\right)\psi_l \nonumber \\
        &+&\bar\psi_h\left(m_h+i\gamma_5\mu_h\tau_3+\epsilon\tau_a\right)\psi_h
         \, , \label{massterms}
\end{eqnarray}
where $\tau_a$ is a Pauli matrix in flavor space.
Should we choose $\tau_a=\tau_3$\cite{PenSV} or $\tau_a\neq\tau_3$\cite{FreR2}?
An advantage of the parallel choice, i.e. $\tau_a=\tau_3$, is that flavors
do not mix, but an important disadvantage is that the fermion determinant
is not real.  Therefore this parallel choice corresponds to fermions that can
be used as valence quarks but not as sea quarks.
The perpendicular choice, $\tau_a\neq\tau_3$, leads to a real fermion
determinant so those quarks can be both sea and valence, but the flavors in the
action mix, as is evident from (\ref{massterms}).  To elucidate the flavor
and isospin structure of this choice, we study the kaon masses in this work.

The notation $(c,s)_\parallel$ and $(c,s)_\perp$ will
respectively denote twisted doublets with the parallel and perpendicular
choices for $\tau_a$ relative to $\tau_3$.  A degenerate doublet is denoted by
$(u,d)_0$.  The three scenarios to be explored are
\begin{tabular}{l}
~~~scenario 1: $(u,d)_0+s$ [$+c$ if desired] \,, \\
~~~scenario 2: $(u,d)_0+(c,s)_\parallel$ \,, \\
~~~scenario 3: $(u,d)_0+(c,s)_\perp$ \,.
\end{tabular}

\section{Kaon mass splittings and lattice artifacts}

\subsection{Chiral perturbation theory}

The chiral Lagrangian for two twisted doublets is\cite{AbdLWW}
\begin{eqnarray}
{\cal L}_{\chi PT} &=& {\cal L}^{(2)} + {\cal L}^{(4)} + \ldots \,, \\
{\cal L}^{(2)} &=& \frac{f^2}{4}{\rm Tr}(D_\mu\Sigma D_\mu\Sigma^\dagger)
         - \frac{f^2}{4}{\rm Tr}(\chi^\dagger\Sigma+\Sigma^\dagger\chi)
         \nonumber \\
     &&  - \frac{f^2}{4}{\rm Tr}(\hat A^\dagger\Sigma+\Sigma^\dagger\hat A)
         \,, \\
{\cal L}^{(4)} &=& -W_8'{\rm Tr}\big[(\hat A^\dagger\Sigma
                   + \Sigma^\dagger\hat A)^2\big] + \ldots \,,
\end{eqnarray}
where
\begin{equation}
\Sigma=\left(\begin{array}{cc} \!e^{i\omega_l\tau_3/2} & 0 \\
                               0 & e^{i\omega_h\tau_3/2}\! \end{array}\right)
       e^{i\Phi/f}
       \left(\begin{array}{cc} \!e^{i\omega_l\tau_3/2} & 0 \\
                               0 & e^{i\omega_h\tau_3/2}\! \end{array}\right)
\end{equation}
for meson matrix $\Phi$.
The quark mass matrix appears as
\begin{equation}
\chi=2B\left(\begin{array}{cc} m_l+i\mu_l\tau_3 & 0 \\
                               0 & m_h+i\mu_h\tau_3+\epsilon_h\tau_a
                               \end{array}\right) \,,
\end{equation}
and the lattice spacing as $\hat A=2W_0a$, using notation familiar from
\cite{RupS,BarRS1,BarRS2}.

To interpret $(u,d)_0 + (c,s)_\perp$, we diagonalize the quark mass matrices,
\begin{eqnarray}
m_l+i\mu_l\tau_3 &=& \sqrt{m_l^2+\mu_l^2}e^{i\tau_3\omega_l} \,, \\
m_h+i\mu_h\tau_3+\epsilon_h\tau_1 &=& e^{i\tau_3\omega_h/2}
                                      Y^\dagger MYe^{i\tau_3\omega_h/2} \,,
\end{eqnarray}
where $Y = \frac{1}{\sqrt{2}}\left(\begin{array}{cc} 1 & 1 \\ -1 & 1
       \end{array}\right)$ is a standard $\pi/4$ rotation, and
the diagonal matrix $M$ is
\begin{equation}
      M = \left(\begin{array}{cc}
      \sqrt{m_h^2+\mu_h^2}+\epsilon_h & 0 \\
      0 & \sqrt{m_h^2+\mu_h^2}-\epsilon_h
      \end{array}\right) \,.
\end{equation}
The resulting mass terms for kaons and D mesons have the form
\begin{equation}\label{LKD}
 \begin{array}{c}
           (\mathbb{K}^+~\bar{\mathbb D}^0~\mathbb{K}^0~{\mathbb D}^-) \\
           \phantom{9} \\ \phantom{9} \\ \phantom{9}\end{array}
           \left(\begin{array}{cccc} m-\delta & \alpha & 0 & 0 \\
           \alpha & m+\delta & 0 & 0 \\
           0 & 0 & m-\delta & -\alpha \\
           0 & 0 & -\alpha & m+\delta \end{array}\right)
           \left(\begin{array}{l}
           \mathbb{K}^- \\ \mathbb{D}^0 \\ \bar{\mathbb K}^0 \\ \mathbb{D}^+
           \end{array}\right)
\end{equation}
where $m$, $\delta$ and $\alpha$ contain various parameters from the chiral
Lagrangian.
Here, mesons are labelled by their identities in the untwisted limit.
They are clearly not mass eigenstates for nonzero twist, $\alpha\neq0$.
To determine meson eigenstates for a general twist, we diagonalize the
matrix in (\ref{LKD}) through a simple change of basis, and obtain
the eigenstates and eigenvalues of Table~\ref{eigentable}.
\begin{table}
\caption{Eigenvalues and eigenstates of the matrix in (\protect\ref{LKD}).
Note that, for general $\theta$, none of these mass eigenstates are isospin
partners.}\label{eigentable}
\begin{tabular}{cc}
\hline
mass eigenstates & mass eigenvalues \\
\hline
$\psi_{u-} = \mathbb{K}^+\cos\theta-\bar\mathbb{D}^0\sin\theta$
& \\
$\psi_{d-} = \mathbb{K}^0\cos\theta+\mathbb{D}^-\sin\theta$
& $m_{u-} = m_{d-} = m-\sqrt{\delta^2+\alpha^2}$ \\
$\psi_{u+} = \mathbb{K}^+\sin\theta+\bar\mathbb{D}^0\cos\theta$
& \\
$\psi_{d+} = \mathbb{K}^0\sin\theta-\mathbb{D}^-\cos\theta$
& $m_{u+} = m_{d+} = m+\sqrt{\delta^2+\alpha^2}$ \\
\hline
\end{tabular}
\end{table}
The rotation (not twist!) angle, $0\leq\theta<\pi/2$, is given by
\begin{equation}
\tan2\theta = \left|\frac{\alpha}{\delta}\right| \,,
\end{equation}
where $\alpha\sim\sin\omega_l\sin\omega_h$
vanishes if either $(u,d)$ or $(c,s)$ is not twisted,
and $\delta\sim\epsilon_h$
vanishes when $(c,s)$ has no explicit mass splitting term.

Is there a correspondence between these eigenstates and the physical mesons?
As a special case, consider $\alpha=0$.  This implies $\theta=0$ and leads to
scenario 1: $(u,d)_0+s+c$.
Quark flavors do not mix; the physical kaons are a (lighter) degenerate
isospin doublet, and the physical D mesons are a (heavier) degenerate isospin
doublet.  This scenario is also discussed in \cite{MunS}.

As a different special case, consider $\delta=0$.  This implies
$\theta=\pi/4$ and leads to $(u,d)_0+(c,s)_0$.  Quark flavors can now be
diagonalized, and mixings are thus avoided.  In fact, this special case then
becomes identical to $(u,d)_0 + (c,s)_\parallel$ with $m_c=m_s$.
The mass-degenerate doublets are $(K^+,D^-)$ and $(K^0,\bar{D}^0)$.
The physical kaons only become degenerate in the continuum limit.

\begin{figure}[th]
\begin{center}
\resizebox{0.30\textwidth}{!}{\includegraphics{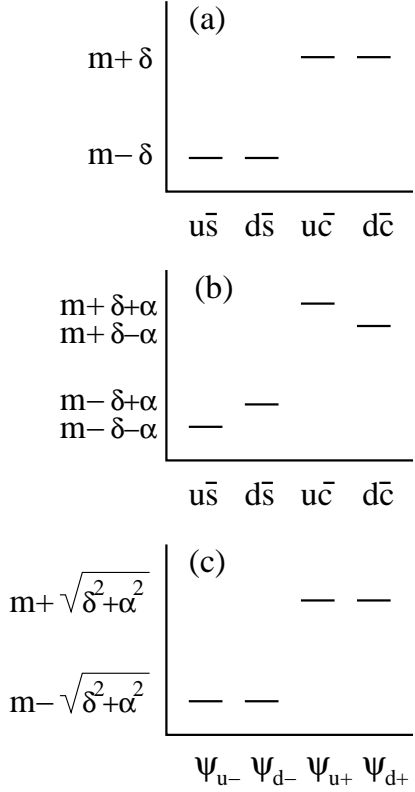}}
\end{center}
\caption{Meson mass eigenstates obtained from three scenarios,
(a) scenario 1: $(u,d)_0 + s + c$,
(b) scenario 2: $(u,d)_0 + (c,s)_\parallel$,
(c) scenario 3: $(u,d)_0 + (c,s)_\perp$.}\label{sketch}
\end{figure}

In general, when $\alpha\neq0$ and $\delta\neq0$, Table~\ref{eigentable}
corresponds to $(u,d)_0 + (c,s)_\perp$.
The light quark doublet has no flavor mixing, but the heavy doublet does.
Interestingly, none of the mass eigenstates form isospin doublets
unless isospin is somehow defined to involve $(c,s)$ as well as $(u,d)$.
Our findings are sketched in Fig.~\ref{sketch}.  Scenario 1 has no isospin
splittings among kaons, and none among D mesons either, but there are $O(a)$
splittings between the kaons and D mesons.
In scenario 2 all splittings begin at $O(a^2)$, but twist artifacts do
cause unphysical $O(a^2)$ isospin splittings between charged and neutral
kaons, and also between charged and neutral D mesons.
In scenario 3, flavors mix.  There are two degenerate
pairs of eigenstates, but the symmetry relating the degenerate states
involves both quark doublets.  Identification of these states with physical
particles only becomes clear upon extrapolation to the continuum limit.

\subsection{Lattice QCD simulations}

According to the chiral Lagrangian discussion above, the splitting
$m^2(K^0)-m^2(K^\pm)$ that arises from twist artifacts in the
$(u,d)_0 + (c,s)_\parallel$ scenario should be, at leading order,
(i) independent of quark mass,
(ii) linear in $a^2$, and
(iii) vanishing as $a^2\to0$.
Quenched numerical simulations for this scenario\cite{AbdLWW,AbdLW} are
consistent with
these expectations, though the vanishing of the mass splitting as $a^2\to0$
is not as clear in the
data as one might have hoped.  Even at the smallest lattice spacing,
$a\approx0.068$ fm, the kaon mass splitting is found to be sizeable:
$m(K^0)-m(K^\pm)\sim50$ MeV.

Ref.~\cite{AbdLWW} also contains quenched results for the $(u,d)_0 + s$
scenario, and these confirm the chiral Lagrangian claim of $m(K^0)=m(K^\pm)$,
i.e. the kaon masses are not split by twist artifacts.

For scenario 3, $(u,d)_0 + (c,s)_\perp$, conventional isospin is only
restored in the continuum limit.  Therefore meson mass eigenstates must be
identified by computing the full meson correlation matrix and then
diagonalizing it numerically.  With quark mass terms
\begin{eqnarray}
{\cal L}&=&\begin{array}{c} (\bar u~\bar d) \\ \phantom{9} \end{array}
           \left(\begin{array}{cc} m_l+i\gamma_5\mu_l & 0 \\
           0 & m_l-i\gamma_5\mu_l \end{array}\right)
           \left(\begin{array}{c} u \\ d \end{array}\right) \nonumber \\
        &+&\begin{array}{c} (\bar q_1~\bar q_2) \\ \phantom{9} \end{array}
           \left(\begin{array}{cc} m_h+i\gamma_5\mu_h & \epsilon_h \\
           \epsilon_h & m_h-i\gamma_5\mu_h \end{array}\right)
           \left(\begin{array}{c} q_1 \\ q_2 \end{array}\right)
\end{eqnarray}
and operators $O_{ui}=\bar u\gamma_5q_i$
and $O_{di}=\bar d\gamma_5q_i$, numerical results give
\begin{eqnarray}
C^{(u)}_{ij} \equiv \left<O_{ui}O^\dagger_{uj}\right> & \Rightarrow &
C^{(u)} = \left(\begin{array}{cc} a & c \\ c & b \end{array}\right) \,, \\
C^{(d)}_{ij} \equiv \left<O_{di}O^\dagger_{dj}\right> & \Rightarrow &
C^{(d)} = \left(\begin{array}{cc} b & c \\ c & a \end{array}\right) \,,
\end{eqnarray}
where $a$, $b$ and $c$ are real-valued.
Diagonalization of $C^{(u)}$ and $C^{(d)}$ provides the meson mass eigenstates.
Simulation results\cite{Far+,Chi+} are consistent with our chiral
Lagrangian discussion.

\section{Summary}

Twisted mass lattice QCD removes $O(a)$ errors and unphysical near-zero modes.
The up and down quarks fit natually into a twisted doublet but the strange
quark has no natural partner.  Three scenarios for including the strange quark
were considered within chiral perturbation theory, and found to have the
following features:

\vspace{2mm}
\noindent
$\bullet$ $(u,d)_0 + s + c$
\begin{itemize}
\item
no unphysical mass splittings within isospin multiplets,
\item
the strange quark does not benefit from twisting,
\item
kaon operators (i.e. twisted+untwisted) require care,
\end{itemize}
$\bullet$ $(u,d)_0 + (c,s)_\parallel$
\begin{itemize}
\item
all quarks benefit from twisting,
\item
suitable for valence quarks but not for sea quarks,
\item
flavors do not mix and isospin is easily managed,
\item
discretization effects appear as kaon mass splittings,
\end{itemize}
$\bullet$ $(u,d)_0 + (c,s)_\perp$
\begin{itemize}
\item
all quarks benefit from twisting,
\item
dynamical simulations can use a single action,
\item
flavors mix; isospin is not straightforward for $a\neq0$.
\end{itemize}
These features have also been observed in explicit numerical simulations, and
may be helpful when choosing which scenario is optimal for a particular
phenomenological application of lattice QCD.

\section*{Acknowledgements}

The authors thank Istv\'an Montvay for helpful communications.
RL is grateful to the organizers of QNP06 for the opportunity to participate
in such an excellent conference.
The work was supported in part by the Natural Sciences and Engineering
Research Council of Canada, the Canada Foundation for Innovation, the Canada
Research Chairs Program, and the Government of Saskatchewan.

\end{document}